\documentclass{ws-ijmpd}
\usepackage[super,compress]{cite}
\begin{document}

\markboth{E.V. Bugaev, P.A. Klimai}
{Primordial black hole constraints for curvaton models}  % 8 words max

\catchline{}{}{}{}{}
%
%%%%%%%%%%%%%%%%%%%%%%%%%%%%%%%%%%%%%%%%%%%%%%%%%%%%%%%%%%%%%%%%%%%%

\title{PRIMORDIAL BLACK HOLE CONSTRAINTS FOR CURVATON MODELS WITH PREDICTED LARGE NON-GAUSSIANITY}

\author{E.V. BUGAEV \footnote{bugaev@pcbai10.inr.ruhep.ru} \; and \; P.A. KLIMAI\footnote{pklimai@gmail.com} }

\address{Institute for Nuclear Research, Russian Academy of Sciences,\\
60th October Anniversary Prospect 7a, 117312 Moscow, Russia }

\maketitle

\begin{history}
\received{...}  % {Day Month Year}
\revised{...}
\end{history}

\begin{abstract}
We consider the early Universe scenario which allows for production of non-Gaussian
curvature perturbations at small scales. We study the peculiarities of a formation
of primordial black holes (PBHs) connected with the non-Gaussianity. In particular,
we show that PBH constraints on the values of curvature perturbation power spectrum
amplitude are strongly dependent on the shape of perturbations and can significantly
(by two orders of magnitude) deviate from the usual Gaussian limit
${\cal P}_\zeta \lesssim 10^{-2}$. We give examples of PBH mass spectra calculations
and PBH constraints for the particular case of the curvaton model.
\end{abstract}


\ccode{PACS numbers: 98.80.Cq }

%\tableofcontents

\section{Introduction}

As is well known, in models of slow-roll inflation with one scalar field the curvature
perturbation originates from the vacuum fluctuations during inflationary expansion, and
these fluctuations lead to practically Gaussian classical curvature perturbations
with an almost flat power spectrum. However, it is well known also that both
these features are not generic in the case of inflationary models with two (or more)
scalar fields: such models can easily predict adiabatic perturbations with,
e.g., a ``blue'' spectrum and these perturbations can be non-Gaussian \cite{Linde:1996gt}.

Possibilities for appearing of non-Gaussian fluctuations in inflationary models
with multiple scalar fields had been discussed long ago
\cite{Salopek:1988qh, Salopek:1991jq, Fan:1992wv}. The time
evolution of the curvature perturbation on superhorizon scales (which
is allowed in multiple-field scenarios \cite{Starobinsky:1986fxa}) implies
that, in principle, a rather large non-Gaussian signal can be generated during inflation.
According to the observational data \cite{Komatsu:2010fb}, the primordial curvature
perturbation is Gaussian with an almost scale-independent power spectrum. So far there is
only a weak indication of possible primordial non-Gaussianity [at $(2-3)\sigma$ level] from
the cosmic microwave background (CMB) temperature
information data (see, e.g., Ref. \refcite{Yadav:2007yy}).
However, non-Gaussianity is expected to become an important probe of both the early and
the late Universe in the coming years \cite{Komatsu:2009kd}.

The second important feature of predictions of two-field models is that these
models can lead to primordial curvature perturbations with blue spectrum (for
scales which are smaller than cosmological ones) and, correspondingly, can predict
the primordial black hole (PBH) production at some time after inflation. In this case, PBHs
become a probe for the non-Gaussianity of cosmological perturbations
\cite{Bullock:1996at, Ivanov:1997ia, PinaAvelino:2005rm, Hidalgo:2007vk}. The results of
PBH searches can be used to constrain the ranges of early Universe model parameters.

There are several types of two-field inflation scenarios in which detectable
non-Gaussianity of the curvature perturbation can be generated: curvaton
models \cite{Mollerach:1989hu, Linde:1996gt, Lyth:2001nq, Moroi:2001ct, Lyth:2006gd},
models with a non-inflaton field causing inhomogeneous reheating \cite{Dvali:2003em, Kofman:2003nx},
curvaton-type models of preheating (see, e.g., Ref. \refcite{Kohri:2009ac} and references therein),
models of waterfall transition that ends the hybrid
inflation \cite{Felder:2000hj, Asaka:2001ez, Copeland:2002ku, GarciaBellido:2002aj}.
In these two-field models the primordial curvature perturbation $\zeta$ has two components:
a contribution of the inflaton (almost Gaussian) and a contribution of the extra field.
This second component is parameterized by the following way \cite{Boubekeur:2005fj}
\begin{equation}
\zeta_\chi({\bf x}) = a \chi({\bf x}) + \chi^2({\bf x}) - \langle \chi^2 \rangle.
\end{equation}
If $a=0$, one has a $\chi^2$-model. Obviously, the quadratic term can't dominate in $\zeta$ on
cosmological scales where CMB data are available. It can, however, be important on smaller scales.

In the present work we study the predictions of the PBH production
and corresponding PBH constraints for the curvaton model.
The potentially large non-Gaussianity in this model is
connected with the fact that the predicted
magnitude of the curvature perturbation is proportional to a square of the
non-inflaton (curvaton) field. The blue spectrum in the curvaton
model is due to, e.g., supergravity effects leading to the large effective mass
of the curvaton \cite{Linde:1996gt}.

The main attention in the present paper is paid to a study of probability distribution
function (PDF) of the curvature perturbation and the shape of the black hole mass
function, with taking into account of the non-Gaussianity. The first general study of
PDF of the curvature perturbation in curvaton model was carried out in Ref. \refcite{Sasaki:2006kq}.

PBH production in curvaton scenario was studied in recent works
\refcite{Kawasaki:2012wr}, \refcite{Firouzjahi:2012iz} (without considering
the non-Gaussian effects). The approximate PBH constraints on the curvature
perturbation power spectrum in the curvaton model were obtained in Ref. \refcite{Lyth:2012yp}.

The plan of the paper is as follows. In Sec. \ref{s2} we study the process
of PBH production in the case when the primordial curvature perturbations
are strongly non-Gaussian. We calculate the PDF function for the $\chi^2$-model
and use it for a calculation of PBH mass spectrum via Press-Schechter mechanism.
In Sec. \ref{sec-curvaton} we discuss the possible
production of PBHs in the curvaton model and the corresponding cosmological constraints that
can be obtained. Our conclusions are given in Sec. \ref{sec-concl}.

\section{PBH mass spectrum in the case of non-Gaussian curvature perturbations}
\label{s2}

\subsection{PDFs for $\chi^2$-model}

Generally, in $\chi^2$-model the connection between curvature perturbation $\zeta$
and the square of the (non-inflaton) scalar field perturbation value $\chi^2$
is
\begin{equation}
\label{zetaPM}
\zeta = \pm A (\chi^2 - \langle \chi^2 \rangle) , \qquad A > 0 .
\end{equation}
The distribution of $\chi$ is assumed to be Gaussian, i.e.,
\begin{equation}
\label{pchiGauss}
p_\chi(\chi) = \frac{1}{\sigma_\chi \sqrt{2\pi}} \; e^{-\frac{\chi^2}{2 \sigma_\chi^2} }, \qquad
\sigma_\chi^2 \equiv\langle \chi^2 \rangle.
\end{equation}
The form of the PDF depends on the sign in front of the right-hand side of Eq. (\ref{zetaPM}).
If the sign is negative, i.e.,
\begin{equation}
\label{zetaAchi2sigchi}
\zeta = - A (\chi^2 - \langle \chi^2 \rangle),
\end{equation}
it is convenient to introduce the notation
\begin{equation}
\zeta_{max} \equiv A \langle \chi^2 \rangle, \qquad \zeta \le \zeta_{max},
\end{equation}
and the PDF is given by the formula \cite{Bugaev:2011wy}
\begin{equation}
\label{pzetaDISTR}
p_\zeta(\zeta) = p_\chi \left| \frac{d\chi}{d\zeta} \right| = \frac{1}{\sqrt{2\pi \zeta_{max}
(\zeta_{max} - \zeta) } } \;
e^{\frac{\zeta - \zeta_{max}}{2\zeta_{max}} },
\end{equation}
which is just a $\chi^2$-distribution with one degree of freedom, with an opposite sign of the argument,
shifted to a value of $\zeta_{max}$.

In a case of the positive sign in Eq. (\ref{zetaPM}) one has, correspondingly,
\begin{equation}
\zeta = A (\chi^2 - \langle \chi^2 \rangle),
\label{zetaAchi2minusAVG}
\end{equation}
and, introducing the notation
\begin{equation}
\zeta_{min} \equiv - A \langle \chi^2 \rangle, \qquad \zeta \ge \zeta_{min},
\end{equation}
one obtains for the PDF the expression
\begin{equation}
p_\zeta (\zeta) = \frac{1}{ \sqrt{2\pi \zeta_{min} (\zeta_{min} - \zeta) } }
 \; e^{\frac{\zeta - \zeta_{min}} {2 \zeta_{min}} }.
 \label{pzeta-curvaton}
\end{equation}
The case when the sign is negative is realized in the model of hybrid inflation
waterfall \cite{Lyth:2010zq, Bugaev:2011qt} and was studied in detail
in Ref. \refcite{Bugaev:2011wy}. In this case,
$\chi$ is the perturbation of the waterfall field.

In our present case, when $\chi$ is the perturbation of the curvaton field,
the sign is positive. The variance of the PDF of the $\zeta$ field distribution is
\begin{equation}
\label{zeta2avg}
\langle \zeta^2 \rangle = \int \limits_{\zeta_{min}}^{\infty} \zeta^2 p_\zeta(\zeta) d \zeta
= 2 \zeta_{min}^2.
\end{equation}
This variance is connected with the curvature perturbation spectrum
${\cal P}_\zeta$ through the expression
\begin{equation}
\label{zeta2avgP}
\langle \zeta^2 \rangle = \sigma_\zeta^2 = \int {\cal P}_\zeta(k) \frac{dk}{k}.
\end{equation}

The distribution function (\ref{pzeta-curvaton}) can be written in the form
\begin{equation}
\label{pzetaFACT}
p_\zeta(\zeta) = \frac{1}{\sigma_\zeta} p\left( \frac{\zeta}{\sigma_\zeta} \right)
 \equiv \frac{1}{\sigma_\zeta} p(\nu), \qquad
 p(\nu) = \frac{1}{\sqrt{1+\sqrt{2}\nu}} e^{-\frac{1}{2}(1+\sqrt{2}\nu)}.
\end{equation}
Here, the ratio $\nu\equiv \zeta/ \sigma_\zeta$ is introduced.
The first central moments of the PDF of the $\zeta$ field % (\ref{zetaAchi2minusAVG})
are given by
\begin{equation}
\langle \zeta \rangle = 0; \quad \langle \zeta^2 \rangle = 2 A^2 \langle \chi^2 \rangle^2; \quad
\langle \zeta^3 \rangle = 8 A^3 \langle \chi^2 \rangle^3; \quad
\langle \zeta^4 \rangle = 60 A^4 \langle \chi^2 \rangle^4,
\label{cms}
\end{equation}
where $\langle \chi^2 \rangle$ is a variance of the $\chi$-field power spectrum,
\begin{equation}
\langle \chi^2 \rangle = \int {\cal P}_\chi(k) \frac{dk}{k}.
\end{equation}
We will use also the central moments of $p(\nu)$ distribution (``reduced central
moments''). They are given by the expression $\langle\zeta^n\rangle / \sigma_\zeta^n$.
In particular, for the skewness and kurtosis one has, respectively,
\begin{equation}
S = \frac{\langle \zeta^3 \rangle}{\sigma_\zeta^3} =
 \frac{\langle \zeta^3 \rangle}{\langle \zeta^2 \rangle^{3/2}}; \quad
D = \frac{\langle \zeta^4 \rangle}{\langle \zeta^2 \rangle^{4/2}}.
\end{equation}

For the following, we need the expression for the PDF of the smoothed curvature
fluctuations, i.e., instead of Eq. (\ref{zetaAchi2minusAVG}) we must use the
smoothed $\zeta$ field,
\begin{equation}
\zeta_R({\bf x}) = A \int d{\bf y} \tilde W(|{\bf x} -{\bf y}| / R ) \chi^2({\bf y}) -
 A \langle \chi^2 \rangle \int d{\bf y} \tilde W(|{\bf x} -{\bf y}| / R ),
\end{equation}
where $\tilde W$ is the window function. We will use the Gaussian form of this function,
in this case its Fourier transform is $W(kR) = \exp(-k^2 R^2 /2)$.
The expressions for the central moments of the corresponding PDF, $p_{\zeta,R}$,
had been derived in Ref. \refcite{Matarrese:2000iz}. The second and third central
moments of $p_{\zeta,R}$ are given by the formulas
\begin{equation}
\langle \zeta_R^2 \rangle = \frac{2}{(2\pi)^6}A^2 \int d{\bf k} d{\bf k'}
P_\chi(k) P_\chi(k') W^2 (|{\bf k} +{\bf k'}| R),
\end{equation}
\begin{eqnarray}
\langle \zeta_R^3 \rangle = \frac{8}{(2\pi)^9}A^3 \int d{\bf k} d{\bf k'} d{\bf k''}
P_\chi(k) P_\chi(k') P_\chi(k'') \times \qquad \qquad\qquad \nonumber \\
\times W(|{\bf k} +{\bf k'}| R) W(|{\bf k'} - {\bf k''}| R) W(|{\bf k} +{\bf k''}| R),
\end{eqnarray}
where $P_\chi(k)$ is the power spectrum of the $\chi$ field,
\begin{equation}
P_\chi(k) = \frac{2\pi^2}{k^3}{\cal P}_\chi(k).
\end{equation}

Now, we suppose, that the PDF of the smoothed $\zeta$ field can be presented
in the factorized form [as in Eq. (\ref{pzetaFACT})],
\begin{equation}
\label{pzr}
p_{\zeta,R} = \frac{1}{\sigma_\zeta(R)} \tilde p\left( \frac {\zeta_R} {\sigma_\zeta(R)} \right)
\equiv \frac{1}{\sigma_\zeta(R)}\tilde p(\nu_R),
\end{equation}
\begin{equation}
\sigma_\zeta(R) = \langle \zeta_R^2 \rangle^{1/2}.
\end{equation}
If Eq. (\ref{pzr}) is approximately correct, the central moments of $\tilde p(\nu_R)$
are weakly dependent on the smoothing scale $R$. It had been shown in Refs. \refcite{Peebles:1998ph},
\refcite{White:1998da}, \refcite{Koyama:1999fc}, \refcite{Seto:2001mg} that
it is really so, for a wide range of scales. In particular, Seto \cite{Seto:2001mg}
showed that, if the spectrum ${\cal P}_\chi$ has a power form, ${\cal P}_\chi\sim k^{t_\chi}$,
the scale dependences of the variance and third central moment of the $\zeta_R$ field
are:
\begin{equation}
\langle \zeta_R^2 \rangle \sim R^{-2 t_\chi}, \quad \langle \zeta_R^3 \rangle \sim R^{-3 t_\chi}.
\end{equation}
Correspondingly, the scale dependence is canceled in the expression for the skewness
parameter, $S_R=\langle \zeta_R^3 \rangle / \langle \zeta_R^2 \rangle^{3/2}$.
Moreover, it appears \cite{Seto:2001mg} (and it is most essential for our case) that,
if $t_\chi \lesssim 1$, the value of $S_R$ is quantitatively close to $S$ [which is equal
to $\sqrt 8$, as follows from Eq. (\ref{cms})]. The analogous check had been performed
in Ref. \refcite{White:1998da} for the next central moment (kurtosis).

It follows from this analysis that the shape of $\nu_R$ distribution is close to the shape of
$\nu$ distribution, and the smoothing effects enter only through the value of the variance,
$\langle \zeta_R^2 \rangle^{1/2} = \sigma_\zeta(R)$.

As we will see in the next Section, in our case the spectrum of the $\chi$-field
has a power form, ${\cal P}_\chi \sim k^{t_\chi}$, and $t_\chi$ is of
order of $1$. Therefore, basing on the above cited works,
we will use Eq. (\ref{pzr}) for the PDF of the smoothed
$\zeta$ field, with $\tilde p(\nu_R)$ having the same form as $p(\nu)$.

%%%%%%%%%%%%%%%%%%%%%%%%%%%%%%%%%%%%%%%%%%%%%%%%%%%%%%%%%%%%%%%%%%%%%%%%%%%%%%%%%%%%%%%%%%%%%

\begin{figure}[!t]
\center %
\includegraphics[width=0.55\columnwidth, trim = 0 5 0 0 ]{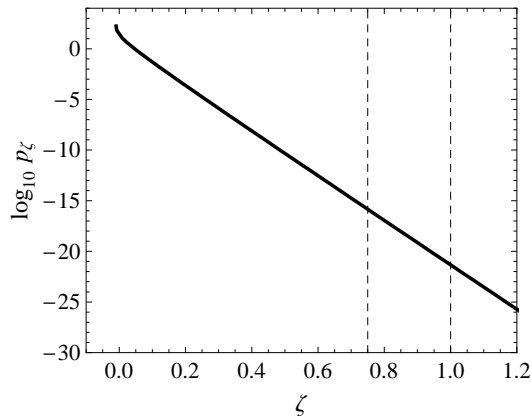}
\caption{ \label{fig-p-zeta-cur}
The form of the distribution (\ref{pzeta-curvaton}) for $\sigma_\zeta^2 = 2\times 10^{-4}$.
Dashed lines show the considered values of $\zeta_c$ ($0.75$ and $1$, see text).
} %
\end{figure}

The form of the distribution (\ref{pzeta-curvaton}) for $\sigma_\zeta^2 = 2\times 10^{-4}$ is
shown in Fig. \ref{fig-p-zeta-cur}. It is seen that in this particular case the probability to
reach $\zeta \sim \zeta_c \sim 1$ (which is, as is well known, required for the
PBH formation) is $\sim 10^{-20}$ or so, i.e., roughly of the same order of magnitude as
PBH constraints on energy density fraction of the Universe contained in PBHs at
the time of their formation, $\beta_{PBH}(M_{BH})$, in PBH mass range
$M_{BH} \sim (10^{10}-10^{20})\;$g \cite{Carr:2009jm}.
It follows from this Figure that the value of ${\cal P}_\zeta(k)\sim 10^{-4}$
is already enough for producing an observable amount of PBHs in this model (this is in
agreement with the estimates of Refs. \refcite{Lyth:2010zq}, \refcite{Lyth:2012yp}).

%%%%%%%%%%%%%%%%%%%%%%%%%%%%%%%%%%%%%%%%%%%%%%%%%%%%%%%%%%%%%
%%%%%%%%%%%%%%%%%%%%%%%%%%%%%%%%%%%%%%%%%%%%%%%%%%%%%%%%%%%%%

\subsection{PBH mass spectrum in Press-Schechter formalism}

For a derivation of PBH mass spectrum and PBH constraints we will use the Press-Schechter
formalism \cite{PS}. We will follow the approach of Refs. \refcite{Lyth:2012yp},
\refcite{Bugaev:2011wy}, and more recent works \refcite{Byrnes:2012yx}, \refcite{Linde:2012bt}
working directly with the curvature perturbation rather than with the density contrast.
In the Press-Schechter formalism, the energy density fraction of the Universe
contained in collapsed objects of initial mass
larger than $M$ is given by
\begin{equation}
\label{PSformalism}
\frac{1}{\rho_i} \int\limits_M^{\infty} \tilde M n(\tilde M) d \tilde M = \int\limits_{\zeta_{c}}^{\infty}
p_{\zeta}(\zeta) d\zeta = P(\zeta>\zeta_{c}; R(M), t_i),
\end{equation}
where function $P$ in right-hand side is the probability that in the region of comoving
size $R$ the smoothed value of $\zeta$ will be larger than the PBH
formation threshold value, $n(M)$ is the mass spectrum of the collapsed objects,
and $\rho_i$ is the initial energy density.
The parameter $\zeta_c$ in Eq. (\ref{PSformalism}) is the threshold of PBH
formation in the radiation-dominated epoch, which
is to be taken from gravitational collapse model. For estimates, in the following we will
use two values: $\zeta_c=0.75$ and $\zeta_c=1$ (corresponding to the PBH formation criterion
in the radiation-dominated epoch: $\delta>\delta_c$, with $\delta_c=1/3$ and $\delta_c=0.45$,
respectively) \cite{Bugaev:2011wy}.

The horizon mass corresponding to the time when fluctuation with initial mass $M$
crosses horizon is (see Ref. \refcite{Bugaev:2008gw})
\begin{equation}
M_h = M_i^{1/3} M^{2/3},
\end{equation}
where $M_i$ is the horizon mass at the moment $t_i$,
\begin{equation}
\label{Mi}
M_i \approx \frac{4\pi}{3} t_i^3 \rho_i,
\end{equation}
and $t_i$ is the time of the start of the radiation era, $\rho_i$ is the energy density
at this time. If the reheating is short, $t_i$
coincides with the time of the end of inflation. In this case, there is a connection between
$M_i$ and the reheating temperature (see, e.g., Ref. \refcite{Bugaev:2008gw}):
\begin{equation}
M_i \approx 0.038 \frac{m_{Pl}^3}{g_*^{1/2} T_{RH}^2}, \qquad g_* \approx 100.
\end{equation}

For simplicity, we will use the approximation that mass of the produced black hole is
proportional to horizon mass, namely,
\begin{equation}
\label{MBH-Mh}
M_{BH} = f_h M_h = f_h M_i^{1/3} M^{2/3},
\end{equation}
where $f_h \approx (1/3)^{1/2} = {\rm const}$.

Using (\ref{PSformalism}) and (\ref{MBH-Mh}), for the PBH number density (mass spectrum)
one obtains \cite{Bugaev:2011wy}
\begin{equation}
\label{nBH}
n_{BH}(M_{BH}) = \left( \frac{4 \pi}{3} \right)^{-1/3}
\left| \frac{\partial P}{\partial R }\right|
 \frac{f_h \rho_i^{2/3} M_i^{1/3} } {a_i M_{BH}^2},
\end{equation}
where $a_i$ is the scale factor at the moment $t_i$.
The derivative ${\partial P}/{\partial R }$ (where $P$ is the function defined in
Eq. (\ref{PSformalism})) is given by the expression
\begin{equation}
\frac{\partial P}{\partial R } = \frac{\zeta_c}{\sigma_\zeta(R)}
 \frac{d \sigma_\zeta(R)}{dR} p_{\zeta,R}(\zeta_c)
\end{equation}
(in derivation of this equation the PDF form of Eq. (\ref{pzr}) was used).

The dependence of the PBH mass spectrum on the curvature perturbation spectrum
${\cal P}_\zeta$ enters just through the factor ${\partial P}/{\partial R }$.

Introducing the PBH {\it formation time}, $t=t_e$ (see Sec. \ref{sec-curvaton}),
we can calculate the energy density fraction of the Universe contained in PBHs,
at the moment $t_e$ (at this moment the horizon mass is equal
to $M_h(t_e) \equiv M_h^f$) \cite{Bugaev:2011wy}:
\begin{eqnarray}
\label{omPBH-beta}
\Omega_{PBH}(M_h^f) \approx \frac{1}{\rho_i} \left( \frac{M_h^f}{M_i} \right)^{1/2} \int n_{BH}(M_{BH}) M_{BH}^2 d \ln M_{BH}
\approx \nonumber \\ \approx
 \frac{1}{\rho_i} \left( \frac{M_h^f}{M_i} \right)^{1/2}
  \left. \left[ n_{BH}(M_{BH}) M_{BH}^2\right] \right|_{ M_{BH} = M_{BH}^{min}} \approx \nonumber \\ \approx
\frac{(M_h^f)^{5/2}}{\rho_i M_i^{1/2}} n_{BH}( M_{BH}) \left. \right|_{ M_{BH} = M_{BH}^{min}} .
\end{eqnarray}
Here, $M_{BH}^{min}$ is the minimum mass of the PBH mass spectrum,  $M_{BH}^{min} = f_h M_h^f$.
It is well known that for an almost monochromatic PBH mass spectrum $\Omega_{PBH}(M_h^f)$ coincides with
the traditionally used parameter $\beta_{PBH}$. Although all PBHs do not form
at the same moment of time, it is convenient to use the combination
$M_i^{-1/2} \rho_i^{-1} M_{BH}^{5/2} n_{BH}(M_{BH})$ to have a feeling of how many PBHs (with mass $\sim M_{BH}$)
actually form, i.e., to use the estimate following from (\ref{omPBH-beta}):
\begin{equation}
\label{beta-estimate}
M_i^{-1/2} \rho_i^{-1} M_{BH}^{5/2} n_{BH}(M_{BH}) \approx \beta_{PBH}.
\end{equation}

\section{PBH constraints in the curvaton model}
\label{sec-curvaton}

Curvaton is an additional to the inflaton scalar field that can be responsible (partly or fully)
for generation of primordial curvature perturbations
\cite{Mollerach:1989hu, Linde:1996gt, Lyth:2001nq, Lyth:2006gd}.
This field can also be a source of PBHs, as discussed in Refs.
\refcite{Kohri:2007qn}, \refcite{Lyth:2010zq}.

\begin{figure}[!b]
\center %
\includegraphics[width=0.55\columnwidth, trim = 0 5 0 0 ]{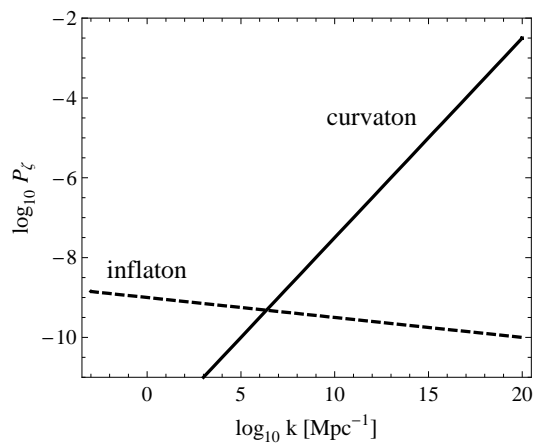}
\caption{ \label{fig-schema}
A sketch that illustrates a relation between curvaton-generated and inflaton-generated
curvature perturbation power spectra for the scenario of PBH production that we consider.
} %
\end{figure}

In this work, we only consider the case of a {\it strong positive tilt}
(the possibility discussed in Refs. \refcite{Linde:1996gt}, \refcite{Lyth:2006gd})
of the curvaton-generated perturbation
power spectrum. At the same time, it is assumed that inflaton is responsible
for generation of perturbations on cosmological scales (see Fig. \ref{fig-schema} for
an illustration).

%%%%%%%%%%%%%%%%%%%%%%%%%%%%%%%%%%%%%%%%%%%%%%%%%%%%%%%%%%%%%
%%%%%%%%%%%%%%%%%%%%%%%%%%%%%%%%%%%%%%%%%%%%%%%%%%%%%%%%%%%%%

The curvaton field generates cosmological perturbations in two stages
\cite{Linde:1996gt, Lyth:2001nq, Lyth:2006gd, Moroi:2001ct}:

(i) Quantum fluctuations of the curvaton during inflation (at time of
horizon exit) become classical, super-horizon perturbations.

(ii) In the radiation-dominated stage, the curvaton starts to oscillate (this happens
at the time when Hubble parameter becomes of order of curvaton's effective mass, $H\sim m$).
The Universe at this stage becomes a mixture of radiation and matter
(the curvaton behaves as a non-relativistic matter in this regime). The pressure
perturbation of this mixture is non-adiabatic and the curvature perturbation is thus
generated. One obtains, approximately, the expression (see, e.g., Ref. \refcite{Enqvist:2005pg})
%\begin{equation}
%\zeta \sim r \delta,
%\label{zeta-r-delta}
%\end{equation}
\begin{equation}
\zeta(t, {\bf x}) = \frac{r \sigma_{\rm osc}'} {2 \sigma_{\rm osc}} \delta\sigma_* +
 \frac{r}{4} \left( \frac{\sigma_{\rm osc}' }{\sigma_{\rm osc}} \right)^2 \delta\sigma_*^2,
\label{zeta-r-delta}
\end{equation}
where $r$ is the density parameter, $r= 4\rho_\sigma/(4\rho_r + 3 \rho_\sigma)$
($\rho_r$ is the energy density of radiation after inflation), $\sigma_{\rm osc}$ is
the value of the curvaton field at the onset of oscillations. The initial value
for the curvaton field, $\sigma_*$, is set by inflation. The derivative in
Eq. (\ref{zeta-r-delta}) is taken with respect to the field value during inflation, $\sigma_*$.
The term containing the second derivative, $\sigma_{\rm osc}''$, is neglected.
It is assumed that $r \ll 1$.

Assuming {\it zero average value} for
the curvaton field (i.e., working with the maximal box \cite{Lyth:2006gd}),
we keep in Eq. (\ref{zeta-r-delta}) only the second term,
\begin{equation}
\zeta(t, {\bf x}) =
 \frac{r}{4} \left( \frac{\sigma_{\rm osc}' }{\sigma_{\rm osc}} \right)^2 \delta\sigma_*^2.
\label{delta-curv}
\end{equation}
In this case we have the $\chi^2$-model (because the perturbations $\delta\sigma_*$ are
assumed to be Gaussian). In notations of Sec. \ref{s2}, one has
$A = \frac{r}{4} \left( \frac{\sigma_{\rm osc}' }{\sigma_{\rm osc}} \right)^2,
\delta\sigma_* = \chi$.
The fluctuations are strongly non-Gaussian which is not forbidden on small scales.

The curvaton-generated curvature perturbation spectrum can be
written \cite{Lyth:2006gd, Bugaev:2012ai} as
(using the Bunch-Davies probability distribution for the perturbations of the
curvaton field \cite{BD, St1982})
\begin{equation}
{\cal P}_{\zeta_\sigma}^{1/2} \sim \frac{2}{3} \Omega_\sigma
\left( \frac{\sigma_{\rm osc}' }{\sigma_{\rm osc}} \right)^2 \frac{1}{\sqrt{t_\sigma}}
\left( \frac{H_*}{2\pi}\right)^2
\left( \frac{k}{k_e} \right)^{t_\sigma},
\label{Pzeta-curvaton}
\end{equation}
where $\Omega_\sigma = \bar \rho_\sigma / \rho \approx r$
is the relative curvaton energy density at the time of its decay (it depends on the
curvaton decay rate) and
\begin{equation}
t_\sigma \cong \frac{2 m_*^2}{3 H_*^2}
\end{equation}
is the tilt of the curvaton field spectrum, ${\cal P}_\sigma\sim k^{t_\sigma}$,
$m_*$ is the effective mass of the curvaton field and $H_*$ is the Hubble parameter
during inflation.

%%%%%%%%%%%%%%%%%%%%%%%%%%%%%%%%%%%%%%%%%%%%%%%%%%%%%%%%%%%%%
\begin{figure}[!bt]
\center %
\includegraphics[width=0.56\columnwidth, trim = 0 5 0 0 ]{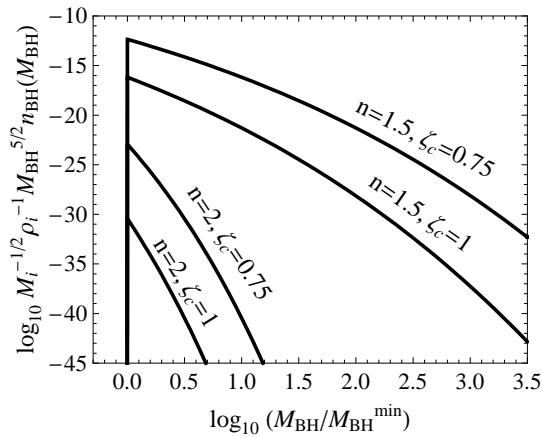}
\caption{ \label{nBH-cur}
Examples of PBH mass spectra calculated for the curvaton model, for several sets of
parameters $n$ and $\zeta_c$. For all curves, ${\cal P}_\zeta ^0 = 4\times 10^{-4}$.
The mass $M_{BH}^{min}=f_h M_h(t_e)$ and $t_e$ is the moment of time
when perturbation with comoving wave number $k_e$ enters horizon.
} %
\end{figure}
%%%%%%%%%%%%%%%%%%%%%%%%%%%%%%%%%%%%%%%%%%%%%%%%%%%%%%%%%%%%%

It is rather natural (see Ref. \refcite{Linde:1996gt} and, e.g., Ref. \refcite{Demozzi:2010aj},
which considers the
models of chaotic inflation in supergravity) that $t_\sigma \sim 2/3$ which
corresponds to a blue perturbation spectrum with the spectral index
\begin{equation}
n = 1 + 2 t_\sigma \sim 7/3
\end{equation}
(such a situation is shown in Fig. \ref{fig-schema}). For the following, we parameterize the
spectrum (\ref{Pzeta-curvaton}) in a simple form
\begin{equation}
{\cal P}_\zeta = {\cal P}_\zeta ^0 \left( \frac{k}{k_e} \right)^{n-1}, \;\; k< k_e,
\label{Pzeta-curvaton-par}
\end{equation}
and will treat ${\cal P}_\zeta ^0$, $n$, and $k_e$ as free parameters. Note that $k_e$
does not, in general, coincide with the comoving wave number corresponding to the end of inflation. Rather,
it is the scale entering horizon at the time when $\zeta$ is created \cite{Lyth:2006gd}
(we assume that it is created instantaneously due to the fast curvaton decay).

Using the Eq. (\ref{nBH}), we can calculate PBH mass distributions that are generated for a
particular set of parameters ($n$, ${\cal P}_\zeta ^0$, etc.) and then compare the resulting
$\beta_{PBH}$ (Eq. (\ref{beta-estimate})) with the known limits
(from, e.g., Ref. \refcite{Carr:2009jm}).
The example of PBH mass spectrum calculation is shown in
Fig. \ref{nBH-cur}.
It is seen from this Figure that PBH
abundances strongly depend on the particular choice of $\zeta_c$.

%%%%%%%%%%%%%%%%%%%%%%%%%%%%%%%%%%%%%%%%%%%%%%%%%%%%%%%%%%%%%
\begin{figure}[!bt]
\center %
\includegraphics[width=0.65\columnwidth, trim = 0 5 0 0 ]{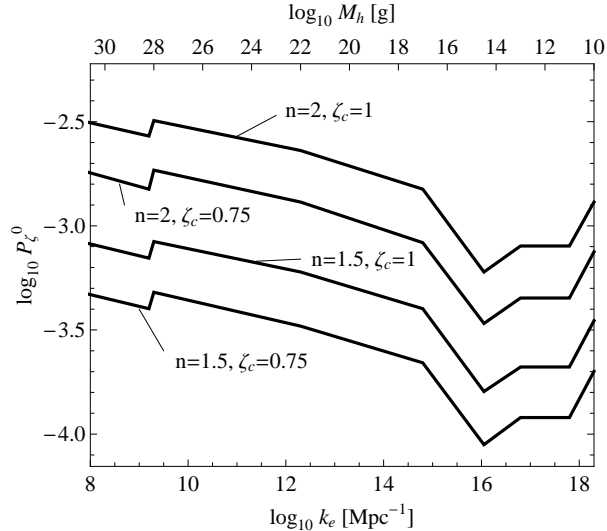}
\caption{ \label{fig-curvaton-limits}
The limits on the maximum value of curvature perturbation power spectrum ${\cal P}_\zeta^0$
from PBH non-observation, for the curvaton model, for several sets of parameters $n$ and $\zeta_c$.
The forbidden regions are above of the corresponding curves.
$M_h$ is the horizon mass at the moment of time when the perturbation with
comoving wave number $k_e$ enters horizon.
The spectrum is parameterized as (\ref{Pzeta-curvaton-par}).
} %
\end{figure}
%%%%%%%%%%%%%%%%%%%%%%%%%%%%%%%%%%%%%%%%%%%%%%%%%%%%%%%%%%%%%

In the calculation of PBH mass spectra we took into account the fact that PBHs
start to form only after the moment of time $t=t_e$ when curvaton decays and $\zeta$
is created \cite{Lyth:2006gd}, and the scale $k_e$ enters horizon just at the same
moment $t_e$. Thus, the minimal PBH mass in our model, taking into account
Eq. (\ref{MBH-Mh}), is
\begin{equation}
M_{BH}^{min} = f_h M_h(t_e).
\end{equation}
This minimal mass corresponds to the vertical line in Fig. \ref{nBH-cur}.

The resulting constraints on parameter ${\cal P}_\zeta ^0$
(for two values of the spectral index $n$) following from data \cite{Carr:2009jm} on PBH
non-observation are shown in Fig. \ref{fig-curvaton-limits}.
Here, the connection between $k_e$ and $M_h$ is given by (see, e.g., Ref. \refcite{Bugaev:2010bb})
\begin{equation}
k_e \approx \frac{2 \times 10^{23}}{\sqrt{M_h / 1 {\rm g}}} {\rm Mpc}^{-1}.
\end{equation}
The obtained constraints can also be transformed into limits
on particular curvaton model's parameters, such as
$\Omega_\sigma$. For example, comparing Fig. \ref{fig-curvaton-limits}
and Eq. (\ref{Pzeta-curvaton}), one obtains, roughly,
\begin{equation}
\Omega_\sigma  \approx  ({\cal P}_\zeta ^0 )^{1/2}  \lesssim 10^{-2},
\end{equation}
which is already a useful constraint. It is a subject of our further study to get more exact
limits for particular parameter sets of the model.

\section{Conclusions}
\label{sec-concl}

Primordial black holes can be used to probe perturbations in our Universe at very small scales,
as well as to study other problems of physics of early stages of the
cosmological evolution. We have considered the PBH
formation from primordial curvature perturbations produced in the curvaton model.
This model predicts the production of
strongly non-Gaussian perturbations, and non-Gaussianity was taken into account in the calculation
of PBH mass spectrum (in Press-Schechter formalism).
Limits on the values of perturbation power spectrum as well as approximate constraints
on inflation model parameters were obtained.
The constraints on the curvature perturbation
spectrum amplitude follow from Fig. \ref{fig-curvaton-limits}.
It had been shown in our previous work \cite{Bugaev:2011wy} that the constraint on ${\cal P}_\zeta$ in the
case when PDF of the curvature perturbation is given by Eq. (\ref{zetaAchi2sigchi}) (this case is realized,
e.g., in the hybrid waterfall model \cite{Bugaev:2011wy}) is very weak,
${\cal P}_\zeta \lesssim 1$. In contrast with this, the corresponding constraint
in the curvaton model is much stronger,
${\cal P}_\zeta \lesssim (10^{-4} - 10^{-2.5})$, depending on the value of the spectral index and
PBH mass. It is important to note that the latter constraints are more
strong than those following from one-field
inflation models, in which ${\cal P}_\zeta \lesssim 10^{-2}$ (see, e.g., Ref. \refcite{Bugaev:2010bb}).
The PBH constraints obtained in this work confirm the estimates given in Refs.
\refcite{Lyth:2010zq}, \refcite{Lyth:2012yp}.

\section*{Acknowledgments}

The work was supported by The Ministry of education and science of Russia, project No. 8525.

%\begin{thebibliography}{000} %for 3 digits

\end{document}